# Generation of THz waves through interaction of the wakefield of two-color laser pulses with magnetized plasma

A. A. Molavi Choobini[1], F. M. Aghamir[2,*]

*Dept. of Physics, University of Tehran, Tehran 14399-55961, Iran.*
*aghamir@ut.ac.ir

**Abstract:** The present study explores radiation in THz spectrum region through the interaction of the wakefield of two-color laser pulses with magnetized plasma. The interaction of the two-color laser with plasma electrons induces transverse nonlinear current in two dimensions, resulting in generation of a wakefield and a forward wave. The investigation revealed that during the non-relativistic regime of laser-plasma interaction, interdependence exists between the electric fields of the forward wave and the wake. Conversely, in the relativistic regime, the dynamic of interaction changes, and plasma electrons are influenced not only by the electric field of the laser pulse but also by relativistic effects like Lorentz contraction, responding to both the electric and magnetic field components. This leads to generation of wake and forward wave radiations. The interplay between various laser and plasma parameters is analyzed, shedding light on the conditions leading to radiation angular distribution patterns in the forward and backward directions. The impact of spatial laser profile, a DC external magnetic field, polarization states, and plasma interaction length on the generated wake and forward wave patterns has been investigated. Through systematic variation of these parameters, the objective is to elucidate the controlled directional features of the resulting fields and radiation patterns.

## 1. Introduction

In recent years, generation of terahertz (THz) radiation has gained significant attention due to the potential applications in various scientific and technological domains [1-3]. The investigation of THz wave radiation has emerged as a fascinating field of study, with profound implications across diverse scientific and technological domains. Hence, understanding the underlying mechanisms and optimizing the THz waves generation is crucial for harnessing their full potential [4, 5]. However, a persistent challenge within this domain pertains to the attainment of narrowband THz sources capable of delivering pulse energies surpassing the threshold of $1\mu J$ [6-9]. The approaches frequently leverage ultrafast laser pulse systems to produce THz radiation, joining diverse mechanisms such as photoconductive antennas and nonlinear optical processes. Nonetheless, a recurring observation is that many of these techniques produce isolated cycles of THz waves characterized by heightened amplitudes [10-12].

In this context, X. Yang et al. investigated high-energy coherent THz radiation emitted from a laser-wakefield scheme [13, 14]. They conducted simulations to analyze the radiation emitted by wide-angle electron beams upon passing through the plasma-vacuum interface. These simulations demonstrated the production of coherent THz radiation with energy levels ranging from $\mu J$ to $mJ$. Yan Peng and colleagues considered the experimental measurement of wakefields in a plasma filament induced by a laser pulse [15]. They introduced a stable method for precise measurement of wakefield amplitude, which influences the intensity and bandwidth of the emitted THz wave. Morgan T. Hibberd and collaborators demonstrated the acceleration of a relativistic electron beam in a THz-driven linear accelerator [16]. The electron beams provide outstanding control over the energy-time phase space of particle bunches, surpassing the capabilities of conventional radiofrequency technology. Alexander Pukhov and his team explored the production of efficient narrow-band THz radiation from electrostatic fields of wakes in non-uniform plasmas [17]. Their study revealed that electrostatic plasma wake can effectively emit radiation at harmonics of plasma frequency, suggesting the possibility of creating a highly efficient, narrow-band, and adjustable source of THz radiation. Employing PIC simulations, THz pulse generation from relativistic laser wakes in axially magnetized plasmas is investigated by C. Tailliez et al. [18]. Their model demonstrated the potential for generating an extra, radially polarized THz field, particularly at high levels of magnetization, a finding corroborated by numerical simulations. In the study conducted by Vivek Sharma et al., a ripple-density plasma was utilized to investigate the axial laser wakefield produced by a laser pulse [19]. They derived analytical formulas to characterize the longitudinal wakefield generated in both homogeneous plasma and plasma with ripple density. J. J. van de Wetering and colleagues studied a plasma-modulated accelerator using the paraxial wave equation and PIC simulations [20]. They demonstrate that the wake amplitude in the accelerator stage can be modulated by adjusting the laser and plasma parameters.

In contrast to the above studies, the current research investigated the interaction of two-color laser pulses with diverse profiles in magnetized uniform plasma. The analysis utilizes induced transverse nonlinear currents in two dimensions, an innovative dual generation of radiation types, leading to production of wakefields and forward waves in THz spectrum region. A comprehensive theoretical model is presented, which offers a nuanced understanding of the conditions that give rise to the wake and forward wave radiation, spanning both relativistic and nonrelativistic regimes. The study reveals that in the non-relativistic regime, the behavior of plasma electrons is predominantly governed by the electric field of the laser pulse. The laser pulse exerts a force on electrons, inducing collective oscillations that generate the forward wave and the wake radiation dependently. In a relativistic regime, the plasma electrons experience substantial relativistic effects, including Lorentz contraction. The contraction alters the perception of space and time for electrons, leading to self-generated magnetic fields arising from the contraction of electrons and their movement. The self-generated magnetic fields interact with the electric field of two-color laser pulses, leading to a decoupling between the electric field of the laser pulse and the wake, independent of each other. Additionally, the interplay between various laser and plasma parameters, such as spatial laser profile, DC external magnetic field, polarization states, and plasma interaction length, is systematically analyzed. This comprehensive analysis sheds light on the conditions leading to specific radiation angular distribution patterns, both in the forward and backward directions. The systematic variation and control of these parameters represent a methodical approach to optimization of THz radiation characteristics in both directions. The article is structured into the following segments: Section II examines the theoretical model of fields of the wake and forward wave generation in both relativistic and nonrelativistic regimes. The next segment showcases the assessment of forward and wake fields, as well as the changes in waveforms of radiation along with the total energy per unit solid angle per unit frequency. Section IV contains a discussion of the outcome and characteristics of forward and wake wave fields. In section V, conclusions are drawn.

## 2. Theoretical Model

Two-color elliptically polarized femtosecond laser pulses with frequencies $\omega_0$ and $2\omega_0$ are considered to be propagating simultaneously along the positive x-direction through magnetized plasma. It is presumed that their polarizations lie in the $y - z$ plane with an arbitrary angle $\theta$ with respect to the $y$-axis. Additionally, an external static magnetic field $\vec{B}_{ext} = B_0 \hat{e}_x$ is applied along the propagation direction. The vector potential of the combined laser pulse can be expressed as:

$$\vec{A}_L(y,z,t) = \vec{A}_1(y,z,t) + \vec{A}_2(y,z,t)$$
$$= A_{01}[Cos\theta \hat{e}_y + Sin\theta \hat{e}_z]e^{i(k_1 x - \omega_0 t)} + A_{02}[Cos\theta \hat{e}_y + Sin\theta \hat{e}_z]e^{i(k_2 x - 2\omega_0 t)} \quad (1)$$

where the wave numbers of the two-color laser pulses are $k_1, k_2$ and $\omega_p$ is plasma frequency. The amplitudes of laser pulses are $A_{0s} = Y(y)Z(z)e^{-t^2/\tau_L^2}$ with $s = 1,2$ and $\tau_L$ is duration of both pulses. The function $Y(y) = Cosh(yb/y_0)e^{-(y/y_0)^q}$ represents the transverse spatial profile of laser pulses, $q$ denotes the index number, $b$ signifies the skew parameter of laser fields, and $y_0$ characterizes the laser pulse width. Through adjustment of these parameters, various types of laser pulse profiles can be accessible. The function $Z(z) = Sin\frac{\pi z}{L}$ denotes the longitudinal spatial profile, where $L$ is the pulse length.

To determine the non-linear current density for evaluation of the electric fields in the relativistic regime, one needs to consider the dynamics of density and the spatial variations of the velocity of plasma electrons through the following equations:

$$\frac{\partial \vec{P}}{\partial t} = c\vec{\nabla}\phi - c\gamma\vec{\nabla}a^2 - c\vec{\nabla}\gamma - \frac{1}{\gamma}\vec{P} \times \vec{\omega}_c \quad (2a)$$

$$\nabla^2 \phi = k_p^2(n_e - 1) \quad (2b)$$

$$\frac{\partial n_e}{\partial t} = -c\vec{\nabla}(n_e \beta) \quad (2c)$$

where $\vec{P}$ is the dimensionless momentum vector normalized to $m_e c$, $\phi$ the normalized scalar potential $(m_e c^2/e)$, $\beta = P/\gamma$ the normalized fluid velocity and $n_e$ is the normalized electron density. The parameter $a_s = eA_{0s}/m_e c^2$ represents the normalized two-color laser vector potential. The asymmetry of a two-color laser pulse brings about electrons to experience various forces in different directions. The phase-dependent force components create asymmetry in the ponderomotive force, imposing electrons oscillation not only in the direction of laser propagation but also transversely. The term $-c\vec{\nabla}\gamma$ in Equation 2a represents the normalized nonlinear ponderomotive force, having both longitudinal (along the propagation direction) and transverse (perpendicular to propagation direction) components. The ponderomotive force pushes electrons away from regions of high laser intensity, leading to charge density modulation and formation of space charges. The laser driving force $(\vec{\nabla}a^2)$ in Eq. (2a) consists of two pulses propagating in the $y$ and $z$-directions with arbitrary amplitudes, the ratio of two pulses are arbitrarily expressed using the transverse wave vectors in the following manner:

$$a^2 = (\vec{a}_1 + \vec{a}_2).(\vec{a}_1 + \vec{a}_2) = a_1^2 + a_2^2 + a_1 a_2 Re(e^{i\Theta_\pm}) \quad (3)$$

where $\Theta_+ = (k_2 - k_1)x - \omega_0 t$ and $\Theta_- = (k_1 - k_2)x + \omega_0 t$, demonstrate two waves, one propagating in the forward direction and the laser wake traveling in opposite direction. Equation 3 magnifies the harmonical oscillation of the density, momentum and scalar potential with the driver's multiple frequencies. Therefore, these quantities along with the relativistic parameter $(\gamma^{-1})$ can be expanded up to the second order and represented based on the harmonics of the driver as follows:

$$P = P_1 Re(e^{i\Theta_\pm}) + P_2 Re(e^{2i\Theta_\pm}) + \cdots \quad (4a)$$

$$\phi = \phi_1 Re(e^{i\Theta_\pm}) + \phi_2 Re(e^{2i\Theta_\pm}) + \cdots \quad (4b)$$

$$n_e = n_{e1} Re(e^{i\Theta_\pm}) + n_{e2} Re(e^{2i\Theta_\pm}) + \cdots \quad (4c)$$

$$\frac{1}{\gamma} = \frac{1}{\gamma_0} - \frac{a_1 a_2}{2\gamma_0^3} Re(e^{i\Theta_\pm}) - \left(\frac{P_1^2}{4\gamma_0^3} - \frac{3a_1^2 a_2^2}{16\gamma_0^5}\right) Re(e^{2i\Theta_\pm}) + \cdots \tag{5}$$

where $\gamma_0 = (1 + a_1^2 + a_2^2)/(1 - \omega_c/\omega)^2$ and $P_1^2 = P_{1x}^2 + P_{1y}^2 + P_{1z}^2$. Using the expansions in Eq. (2) along with continuity equation leads to a set of algebraic equations from which different orders of electron density and momentum can be obtained through an iterative process. The initial iteration yields:

$$-\omega_0 P_{1y} = -\frac{\omega_p^2}{c(k_2 - k_1)} n_{1e}^+ - c(k_2 - k_1)\frac{a_1 a_2}{2\gamma_0} - \frac{\omega_c \omega_0}{c(k_2 - k_1)} n_{1e}^+ + \omega_c P_{1z} \tag{6a}$$

$$+\omega_0 P_{1z} = +\frac{\omega_p^2}{c(k_2 - k_1)} n_{1e}^- + c(k_2 - k_1)\frac{a_1 a_2}{2\gamma_0} + \frac{\omega_c \omega_0}{c(k_2 - k_1)} n_{1e}^- - \omega_c P_{1y} \tag{6b}$$

Therefore, the first-order modified plasma electrons density profile is:

$$n_{1e}^\pm = \pm \frac{\left[\frac{ck_p^2(\omega_0 - \omega_c)}{k_2 - k_1} - \frac{\omega_0(\omega_0^2 + \omega_c^2)\omega_c \omega_0 + \omega_p^2}{\omega_0^2 - \omega_c^2}\right]\left[c(k_2 - k_1)\frac{a_{01} a_{02}}{2\gamma_0}\left(\frac{\omega_0(\omega_0^2 + \omega_c^2)}{\omega_0^2 - \omega_c^2}(\omega_c - \omega_0)\right)\right]}{\left(\frac{c\omega_0 k_p^2}{k_2 - k_1} - \frac{\omega_0(\omega_0^2 + \omega_c^2)\omega_c \omega_0 + \omega_p^2}{\omega_0^2 - \omega_c^2}\right)^2 - \left(\frac{c\omega_0 k_p^2}{k_2 - k_1}\right)^2} e^{i\Theta_\pm} = \zeta^\pm e^{i\Theta_\pm}$$

(7)

Proceeding with the second iteration:

$$P_{2y} = \frac{\omega_0 \gamma_0}{c(k_2 - k_1)}\left[n_{2e}^+ + \left(\frac{a_1^2 a_2^2}{4\gamma_0^4} - 1\right)\frac{\omega_0 \gamma_0}{c(k_2 - k_1)}(n_{1e}^+)^2\right] \tag{8a}$$

$$P_{2z} = \frac{\omega_0 \gamma_0}{c(k_2 - k_1)}\left[n_{2e}^- + \left(\frac{a_1^2 a_2^2}{4\gamma_0^4} - 1\right)\frac{\omega_0 \gamma_0}{c(k_2 - k_1)}(n_{1e}^-)^2\right] \tag{8b}$$

The modified plasma electron density profile in the second order will be:

$$n_{2e}^\pm = \pm \frac{c^2(k_2 - k_1)^2 + 3\omega_0^2(\zeta^\pm)^2 - 2\omega_0^2 \zeta^\pm}{\left(4\omega_0^2 - \omega_c \omega_0 - \frac{\omega_0(\omega_0^2 + \omega_c^2)\omega_c \omega_0 + \omega_p^2}{\omega_0^2 - \omega_c^2} \frac{\omega_p^2}{c(k_2 - k_1)} \pm \frac{\omega_p^2}{\gamma_0}\right)} \left(\frac{a_1 a_2}{2\gamma_0^2}\right)^2 e^{2i\Theta_\pm} \tag{9}$$

Upon insertion of the above expressions for electron momentum and the modified electron density profile into $\vec{J} = e n_e \vec{P}/\gamma m_e$, the resulting transverse nonlinear current density is obtained ensuring appropriate phase matching:

$$\vec{J} = \frac{e}{m_e}\left[(n_{1e}^+)^* P_{2y} + n_{2e}^+ P_{1y}^*\right]\hat{e}_y + \frac{e}{m_e}\left[(n_{1e}^-)^* P_{2z} + n_{2e}^- P_{1z}^*\right]\hat{e}_z \tag{10}$$

$$= J_0^+ e^{i((k_2 - k_1)x - \omega_0 t)}\hat{e}_y + J_0^- e^{i((k_1 - k_2)x + \omega_0 t)}\hat{e}_z$$

where $J_0^\pm$ is:

$$J_0^\pm = \pm \left(\frac{a_1 a_2}{2\gamma_0^2}\right)^3$$

$$\times \left[\frac{c^2(k_2 - k_1)^2 + (\gamma_0 \omega_0^2 - \omega_c \omega_0 + \omega_p^2)\zeta^\pm}{c\omega_0(k_2 - k_1)} \frac{c^2(k_2 - k_1)^2 + 3\omega_0^2(\zeta^\pm)^2 - 2\omega_0^2 \zeta^\pm}{\left(4\omega_0^2 - \omega_c \omega_0 - \frac{\omega_0(\omega_0^2 + \omega_c^2)\omega_c \omega_0 + \omega_p^2}{\omega_0^2 - \omega_c^2} \frac{\omega_p^2}{c(k_2 - k_1)} \pm \frac{\omega_p^2}{\gamma_0}\right)} \pm \left(\frac{a_1^2 a_2^2}{4\gamma_0^4} - 1\right)\frac{\omega_0 \gamma_0}{c(k_2 - k_1)^2}(\zeta^\pm)^3\right] \tag{11}$$

The angular distribution of the generated electric fields of the forward wave and the wake significantly relies on the electromagnetic radiation fields. Utilizing the nonlinear current density as expressed in Eq. (10), the fields can be evaluated for an interaction length $L$ and radius $r_p$. The retarded vector potential can be written as:

$$\vec{A}_\pm(\vec{r},t) = \frac{\mu_0}{4\pi} \int \frac{\vec{J}_\pm(r', \ t-|r-r'|/c)}{|r-r'|} d^3x' \quad (12)$$

Applying the far field approximation $|r - r'| \approx r - x' \cos\theta$ in Eq. (12), where $\theta$ representing the angle between position vector $r$ and $x$-axis (Fig. 1), the electric fields of the forward wave and the wake can be verified:

$$\vec{E}_+(\vec{r},t) = -i\omega_0 A_+ = \frac{\mu_0 c J_0^+ r_p^2}{4r} \frac{e^{i\frac{\omega_0 L}{c}\left(1+\frac{3\omega p/\sqrt{\gamma_0}}{4\omega_0}-\cos\theta\right)}-1}{\left(1+\frac{3\omega p/\sqrt{\gamma_0}}{4\omega_0}-\cos\theta\right)} e^{-i\omega_0\left(t-\frac{r}{c}\right)}\hat{e}_y \quad (13a)$$

$$\vec{E}_-(\vec{r},t) = +i\omega_0 A_- = \frac{\mu_0 c J_0^- r_p^2}{4r} \frac{e^{i\frac{\omega_T L}{c}\left(1-\frac{3\omega p/\sqrt{\gamma_0}}{4\omega_0}-\cos\theta\right)}-1}{\left(1-\frac{3\omega p/\sqrt{\gamma_0}}{4\omega_0}-\cos\theta\right)} e^{+i\omega_0\left(t-\frac{r}{c}\right)}\hat{e}_z \quad (13b)$$

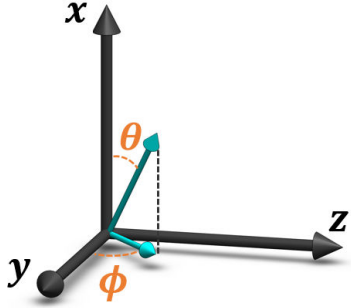

Fig. 1. Schematic of coordinate system used for evaluation of retarded vector potential of forward and backward (wake) waves in far-field approximation.

The time-averaged Poynting vectors of the fields can be written as:

$$\vec{S}_{avg}^+ = \frac{\mu_0 c r_p^4 (J_0^+)^2}{32 r^2} \frac{\left[e^{i\frac{\omega_0 L}{c}\left(1+\frac{3\omega p/\sqrt{\gamma_0}}{4\omega_0}-\cos\theta\right)}-1\right]^2}{\left(1+\frac{3\omega p/\sqrt{\gamma_0}}{4\omega_0}-\cos\theta\right)^2} [1-\sin^2\theta \cos^2\varphi]\hat{e}_r \quad (14a)$$

$$\vec{S}_{avg}^- = \frac{\mu_0 c r_p^4 (J_0^-)^2}{32 r^2} \frac{\left[e^{i\frac{\omega_0 L}{c}\left(1-\frac{3\omega p/\sqrt{\gamma_0}}{4\omega_0}-\cos\theta\right)}-1\right]^2}{\left(1-\frac{3\omega p/\sqrt{\gamma_0}}{4\omega_0}-\cos\theta\right)^2} [1-\sin^2\theta \sin^2\varphi]\hat{e}_r \quad (14b)$$

And the total emitted energy of the fields can be computed through:

$$W^\pm = r^2 \int \vec{S}_{avg}^\pm \cdot \hat{e}_r dt \int d\Omega \quad (15)$$

where the surface normal vector is $\hat{e}_r = \cos\theta\, \hat{e}_x + \sin\theta \cos\varphi\, \hat{e}_y + \sin\theta \sin\varphi\, \hat{e}_z$ and $\varphi$ represents the angle between the position vector $r$ and $y$-axis. Employing the Fourier transform in conjunction with far-field approximation, it becomes possible to derive the total energy per frequency per solid angle:

$$\frac{\partial^2 W_+}{\partial \omega \partial \Omega} = \frac{\mu_0 \tau_L c^2 r_p^4 (J_0^+)^2}{16\sqrt{\pi}} \frac{Sin^2\left[\frac{L\omega}{2c}\left(1+\frac{3\omega_p/\sqrt{\gamma_0}}{4\omega_0}-Cos\,\theta\right)\right]}{\left(1+\frac{3\omega_p/\sqrt{\gamma_0}}{4\omega_0}-Cos\,\theta\right)^2} e^{\frac{-\omega^2 \tau_L^2}{4}} [1 - Sin^2\theta\,Cos^2\varphi] \qquad (16a)$$

$$\frac{\partial^2 W_-}{\partial \omega \partial \Omega} = \frac{\mu_0 \tau_L c^2 r_p^4 (J_0^-)^2}{16\sqrt{\pi}} \frac{Sin^2\left[\frac{L\omega}{2c}\left(1+\frac{3\omega_p/\sqrt{\gamma_0}}{4\omega_0}-Cos\,\theta\right)\right]}{\left(1+\frac{3\omega_p/\sqrt{\gamma_0}}{4\omega_0}-Cos\,\theta\right)^2} e^{\frac{-\omega^2 \tau_L^2}{4}} [1 - Sin^2\theta\,Sin^2\varphi] \qquad (16b)$$

where $\tau_L$ represents the laser pulse duration. Equations (16a, b) illustrate the angular distribution of radiated power for both the forward wave and the wake radiation. In the non-relativistic regime, the interaction of plasma electrons with the two-color laser pulses leads to the first-order quiver velocity, which has components along both $y$ and $z$ directions:

$$V_y^{(1)} = a_1 b_1 e^{i(k_1 x - \omega_0 t)} + a_2 b_2 e^{i(k_2 x - 2\omega_0 t)} \qquad (17a)$$

$$V_z^{(1)} = a_1 b_3 e^{i(k_1 x - \omega_0 t)} + a_2 b_4 e^{i(k_2 x - 2\omega_0 t)} \qquad (17b)$$

where the parameters of $b_1$, $b_2$, $b_3$ and $b_4$ are given as:

$$b_{1,3} = \frac{e}{s m_e \omega_0} \frac{[(v_e - si\omega_0)Cos\theta - \omega_c Sin\theta]}{(v_e - si\omega_0)^2 + \omega_c^2} \qquad (18a)$$

$$b_{2,4} = \frac{e}{s m_e \omega_0} \frac{[(v_e - si\omega_0)Sin\theta - \omega_c Cos\theta]}{(v_e - si\omega_0)^2 + \omega_c^2} \qquad (18b)$$

with $s = 1,2$, and $\omega_c$ is the cyclotron frequency. The second order perturbative expansion of Eq. (2) yields,

$$\frac{\partial \vec{V}^{(2)}}{\partial t} + v_c \vec{V}^{(2)} = \frac{-e}{m_e} \vec{E}_L^{(2)} - \frac{1}{2}\vec{\nabla}\left(\vec{V}^{(1)} \cdot \vec{V}^{(1)}\right) - \vec{V}^{(2)} \times \vec{\omega}_c \qquad (19)$$

The terms $\vec{V}^{(1)} \times \vec{B}^{(1)}$ and $\vec{V}^{(1)} \times (\vec{\nabla} \times \vec{V}^{(1)})$ representing vortex motion do not show up and cancel each other out. Substituting for the first-order velocity components from Eq. 17 into Eq. 19 yields the second-order components of the Lorentz force. Defining the independent variables $\xi(= z - ct)$ and $\tau(= t)$ and applying quasistatic approximation to the time-dependent Maxwell's equations, a pair of coupled second order differential equations for the electric fields of the wake and the forward wave is obtained:

$$\frac{\partial^2 E_+}{\partial \xi^2} + k_p^2 E_+ = \frac{b_1^2 + b_2^2}{4\pi e n_0}\frac{\partial a_1^2}{\partial y} + \frac{b_3^2 + b_4^2}{4\pi e n_0}\frac{\partial a_2^2}{\partial y}$$

$$+ \frac{b_1^* b_2 + b_3^* b_4}{4\pi e n_0}\left(\frac{\partial a_1^*}{\partial y} a_2 + a_1^* \frac{\partial a_2}{\partial y}\right) Cos k_p \xi + \omega_c \frac{\partial E_-}{\partial \xi} \qquad (20a)$$

and

$$\frac{\partial^2 E_-}{\partial \xi^2} + k_p^2 E_- = \frac{b_1^2 + b_2^2}{8\pi e n_0} a_1 \frac{\partial a_1}{\partial \xi} + \frac{b_3^2 + b_4^2}{8\pi e n_0} a_2 \frac{\partial a_2}{\partial \xi}$$

$$+ \frac{b_1^* b_2 + b_3^* b_4}{8\pi e n_0}\left(\frac{\partial a_1^*}{\partial \xi} a_2 + a_1^* \frac{\partial a_2}{\partial \xi}\right) Cos k_p \xi + \frac{b_1^* b_2 + b_3^* b_4}{4\pi e n_0} Sin k_p \xi - \omega_c \frac{\partial E_+}{\partial \xi} \qquad (20b)$$

These equations can be evaluated numerically. The subsequent section provides an in-depth discussion of the results derived from numerical computations.

## 3. Results and Discussion

The interaction between two-color laser pulses and magnetized plasma, spanning relativistic and non-relativistic regimes, is investigated. The emphasis has been placed on understanding the behavior of the wakefield, as well as the forward waves within this context. The simultaneous propagation of the forward and backward THz waves has significant potential and provides insight into the fundamental properties of wave propagation in different media. The bidirectional capability allows for simultaneous transmission and reception of data, improving the overall security and robustness of the communication system. By collecting THz signals from both directions, it is possible to reconstruct a complete 3D image of an object. This is particularly useful in medical imaging, where detailed internal structures need to be visualized, and in security screening, where hidden objects must be detected. To fully harness the benefits of simultaneous forward and backward THz radiation, it is essential to achieve precise control and optimization of the laser and plasma parameters. To implement the theoretical results presented in previous section and for simulations runs, the experimental parameters listed in references [21-23] were considered. A short laser pulse ($> 27 fs$) at $\lambda = 800\ nm$ wavelength with energy parameter $< 2.7 J$ was taken into consideration. The laser focal length was 1.5m with focal spot size $r_L = 22 \mu m$ at FWHM, which provided a peak intensity $I_0 = 5.2 \times 10^{18}\ W/cm^2$. For the sake of comparison, the variations of the angular distribution function and the electric field for wakefield, as well as the forward waves, are presented in relativistic and non-relativistic regimes at the same time respectively, to avoid redundancy.

Figure 2a shows variations of the magnitude of normalized electric fields of forward and backward directed wake as a function of normalized distance. The contours of the normalized electric fields are presented in Figures 2b and 2c. The simulations were performed by COMSOL Multiphysics software. As the Figures show, the normalized electric fields of both forward and backward (wake) waves exhibit a distinctive pattern, namely, an initial rapid surge succeeded by a gradual decline as the distance from the source increases. This behavior suggests that as the waves propagate, they experience energy dissipation and spread. In the initial stage, the fields showcase a pronounced surge in their intensity. This rapid increase suggests an initial concentration of energy, indicative of a tightly focused structure. The sharp electric field intensity during this phase, implies a localized and intense interaction with the surrounding environment. However, as the fields of the wake and forward waves progress in their propagation, a noticeable shift occurs. The subsequent gradual decrease in electric field intensity indicates a transformation in the spatial distribution of energy. This evolution implies that, over distance, the energy associated with the fields disperses, spreading out over a larger spatial domain. The initial focus gives way to a broader distribution, signifying a substantial energy dissipation and spatial spreading. Due to the dynamics of plasma-based wake and forward waves, a notable phenomenon known as wave breaking occurs, followed by a consequential process called phase mixing. Phase mixing is a dynamic mechanism wherein the phases of different components within the forward wave and wakefield evolve at disparate rates. This asynchronous evolution results in a profound redistribution of energy, occurring not only in the spatial dimensions but also in the velocity space. The phase mixing instigates a broadening of the spatial structure of both fields. The once well-defined and compact structures transform, extending over a larger spatial domain. As the phases of different components evolve asynchronously, energy is transferred to modes with varying frequencies and spatial profiles. This intricate energy redistribution process extends beyond the original modal components, leading to the dissipation of energy into a spectrum of higher-order modes.

The effect of various harmonics of laser pulses on the magnitude of normalized electric fields of (a) forward and (b) wake radiations as a function of normalized independent variable ($\xi$) over interaction length for diverse laser profiles in non-relativistic regime is drawn in Figure

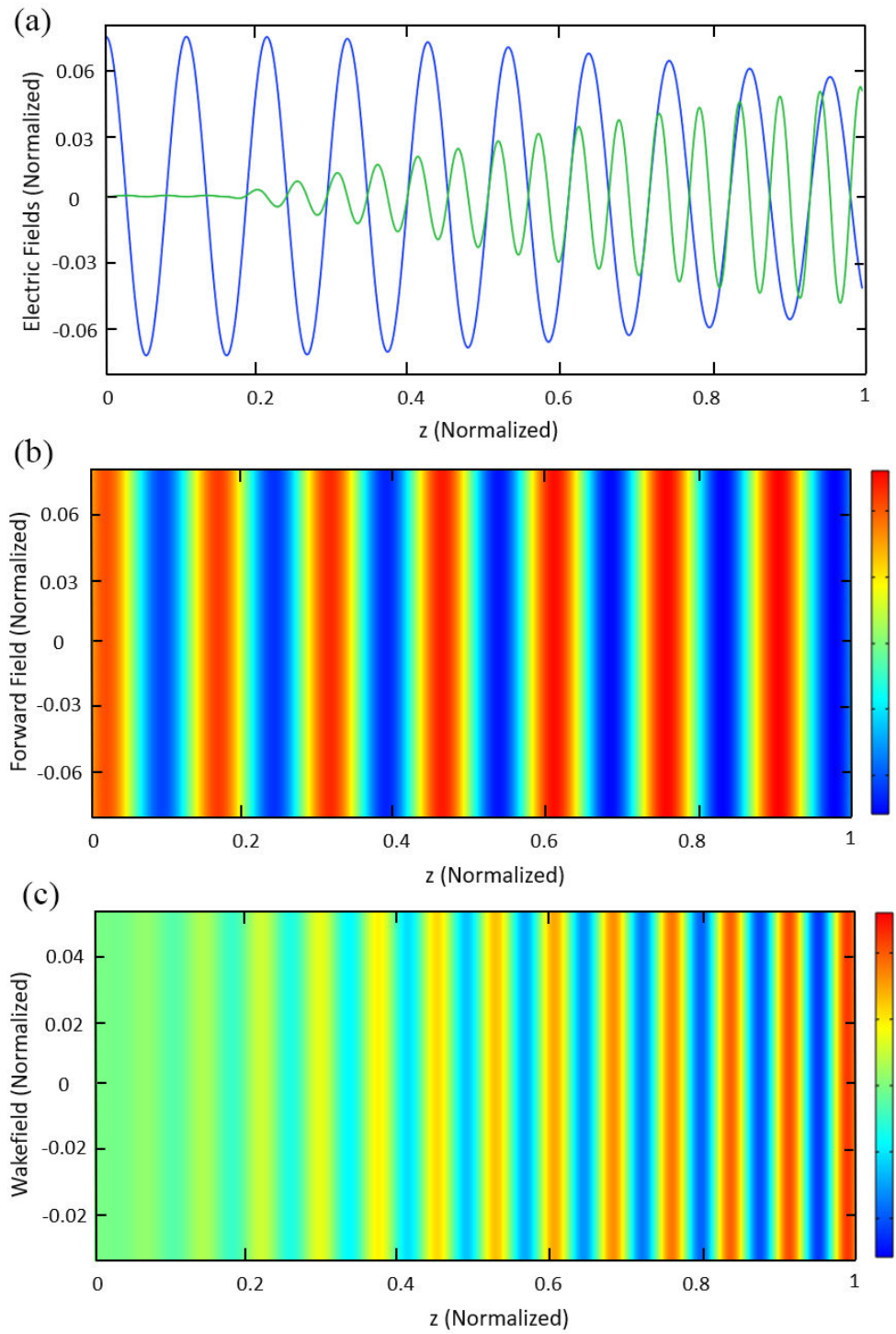

Fig. 2. (a) Variations of normalized electric fields magnitudes of the wake and forward waves as a function of normalized distance, along with contour plots of the normalized electric fields ((b) forward waves and (c) wake waves) as a function of normalized distance.

3. Higher harmonics in laser pulses modify the plasma wave structure and influence resonance conditions for wave-particle interaction through processes like harmonic generation, frequency mixing, and enhanced nonlinear effects. When these higher harmonics interact with plasma, they exert forces on plasma electrons in a nonlinear fashion, leading to enhanced interaction with plasma particles. The resonant interaction between the laser and plasma becomes a multi-frequency phenomenon, with different harmonics contributing to the overall dynamics. These additional harmonic frequencies can lead to the creation of new plasma waves. This process generates sidebands around the primary plasma frequency, affecting the resonance condition. The presence of higher harmonics alters the effective plasma frequency experienced by electrons, affecting the synchronization between the laser and plasma waves. This modification leads to enhancement in the energy transfer and acceleration processes during wave-particle interactions. Therefore, an increase in the harmonics of laser pulses boosts the magnitude of normalized electric fields of forward and wake radiations. In addition, the laser pulse profile influences the plasma's spatial and temporal intensity distribution. This, in turn, affects the dynamics of electron acceleration, leading to variations in the forward wave and the wake structure in THz zone. The interaction between the laser electric field and the charged particles in the plasma is modulated by the chosen profile, determining the efficiency and pattern of electron acceleration. When interacting with plasma, the Gaussian profile spatial pattern orchestrates a relatively uniform acceleration of electrons, contributing to a more structured and symmetric field of wake. Conversely, profile variation introduces spatial complexities, with potentially sharper or flatter intensity peaks. This intricate spatial distribution induces non-uniformities in the acceleration of electrons. The accelerated electron population, influenced by the laser profile, becomes a source of electromagnetic radiation, and the characteristics of this radiation, including its frequency spectrum and angular distribution, reflect the intricacies of the wake it originates from. Figure 4 depicts the various patterns of angular distributions of the wake and forward wave for different laser profiles in a relativistic regime. The Figure clarifies the variations of the angular distributions, showcasing how alterations in laser profiles induce diverse patterns in the emitted radiation for both the fields of forward and wake radiations. These variations include differences in the intensity and width of the emitted radiation at different angles. The Figure shows that the spatial characteristics of the laser pulse determine how efficiently energy is transferred to the plasma and subsequently radiated. Different profiles result in distinct energy distributions across angular regions, leading to variations in intensity. Variations in laser profiles introduce changes in the spatial extent over which the radiation is distributed. This affects the width of the angular patterns observed in the fields of forward and wake radiation emission. The experimental findings by Yan Peng et al. confirm the observed results related to the normalized fields of the wake in induced plasma by a single-color ultrafast laser pulse [24]. The findings of present study also agree with the results of Danni Ma et al. on the normalized THz waves electric field initiated by two-color laser pulse [25].

The fields of forward and wake radiation rely on the generation of plasma waves driven by an external energy source, such as a laser or a particle beam. These waves induce alternating electric fields within the plasma and serve as a platform for either accelerating charged particles or emitting THz radiation. The effect of various external magnetic fields on variations of magnitude of normalized electric fields of (a) forward and (b) wake radiations as a function of normalized independent variable ($\xi$) over interaction length in a non-relativistic regime is presented in Fig 5. As the Figure shows, alterations in the electron motion and density distribution result in modifications of forward wave and wakefield frequencies along with their amplitudes. This is due to the magnetic field, affecting the electron motion and influencing the plasma characteristic frequencies. Changes in electron motion and density distribution affect the conditions for resonance, prompting which modes are amplified in all laser profiles. The nonlinear current is linked to the density modulation induced by the external energy source and is modified by the magnetic field. As a result, the strength of the electric fields in the wake, or the intensity of the emitted radiation in the forward direction in the THz zone, is enhanced by increasing the external magnetic field. These outcomes are confirmed by C. Tailliez et al. through experiments related to the normalized field of plasma induced wake [26]. Figure 6 shows the impact of the external magnetic field on the angular distributions of the forward wave and the wakefield in THz zone for Gaussian profile and elliptical polarization state in the relativistic regime. The Figure indicates that the inclusion of an external magnetic field is not merely a nuanced detail but a decisive factor influencing the dynamics of the system and modulating the behavior of charged particles within the plasma. The external magnetic field amplifies the transverse motion of electrons, leading to an intensified ponderomotive force exerted by the laser field on charged particles. The larger ponderomotive force, facilitated by the external magnetic field, plays a pivotal role in amplifying the fields. The areas where wakefields show up serve as regions of alternating electric fields that can accelerate charged particles. Simultaneously, the rise of ponderomotive force conduces to enhanced emission of forward and wake waves, which are associated with oscillations of plasma waves. Furthermore, the rise of the external magnetic field does not coincide with the appearance of side lobes in the angular distributions. Side lobes typically represent additional peaks or features in the angular distribution. The absence of side lobes with increasing magnetic field strength signifies the profound influence of the magnetic field on the emitted radiations. Thus, the strength and orientation of the external magnetic field can be tuned to control the characteristics of the forward wave and the wavefield.

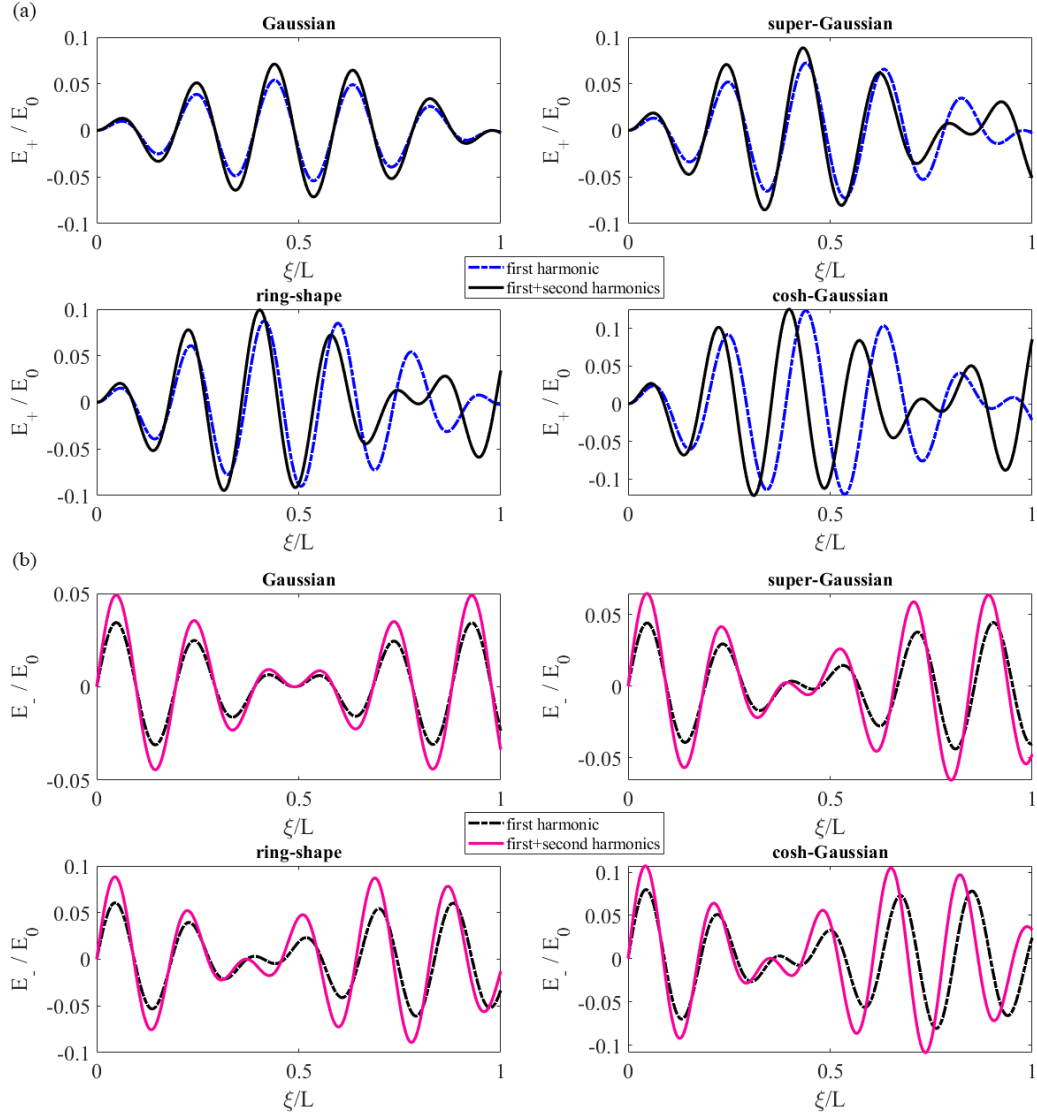

Fig. 3. Effect of various harmonics of laser pulses on the magnitude of normalized electric fields for elliptical polarization (a) forward and (b) wake radiations as a function of normalized independent variable $\xi (= z - ct)$ over plasma interaction length ($L$) for different laser profiles in non-relativistic regime, $\lambda = 800\ nm$, $L = 10\text{mm}$, $B_{ext} = 100\ T$.

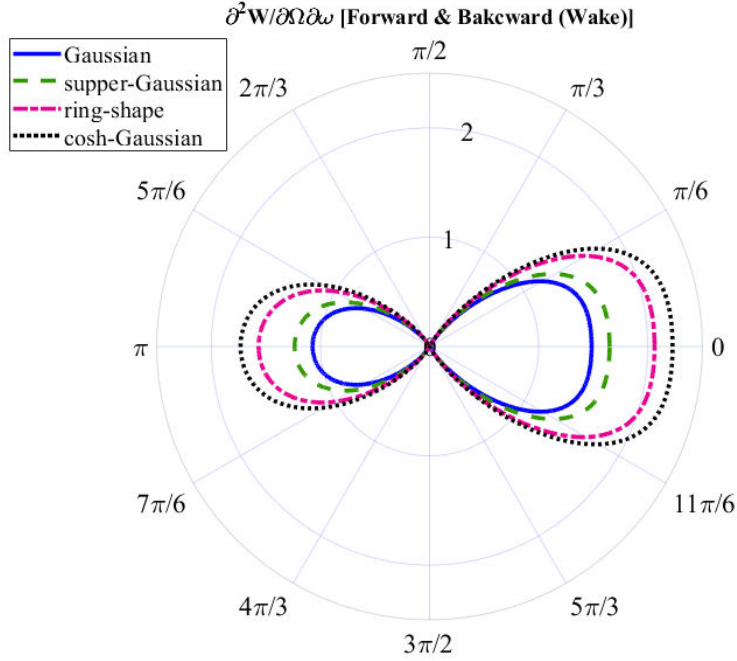

Fig. 4. Patterns of angular distributions of (a) forward and (b) wake radiations for elliptical polarization and different laser profiles in relativistic regime, $\lambda = 800\ nm$, $L = 10\text{mm}$, $B_{ext} = 100\ T$.

The impact of laser pulse polarization states and profiles on fields of the wake and forward wave radiation in the non-relativistic regime is highly dependent on specifics of experimental setup. Understanding these relationships requires detailed numerical simulations tailored to the specific laser and plasma conditions. Figure 7 displays the role of various polarization states of laser pulses on the magnitude of normalized electric fields of (a) forward emission and (b) wake radiation as a function of the normalized independent variable ($\xi$) over interaction length for different laser profiles in non-relativistic regime. The Figure reveals insight into the effects of different polarization states across all laser pulse profiles. In the case of elliptical polarization, the electric field exhibits a higher magnitude compared to circular polarization. The period of electric field oscillation is shorter, signifying a rapid oscillation. The elliptical nature of polarization leads to asymmetric field of structure. The varying strengths and orientations of the electric field components create forward field, as well as wakefield that are not symmetric, potentially introducing novel effects in the acceleration of plasma electrons. The plasma electrons will experience different acceleration strengths and directions depending on their initial positions and velocities relative to the laser pulse's electric field orientation. This directional variability leads to non-uniform electron energy distribution. Different regions of plasma experience distinct acceleration forces and trajectories, leading to spatial variations in electron density and as a result, various nonlinear current densities. The plot of angular distributions of both the forward wave and wakefield for different polarization states of laser pulses in the relativistic regime is depicted in Fig 8. According to the Figure, for the elliptical polarization, the asymmetry of the electric field components causes an effective transfer of energy and momentum to the plasma electrons. This turns into wider distribution function in both forward and wake modes compared to circular polarization states.

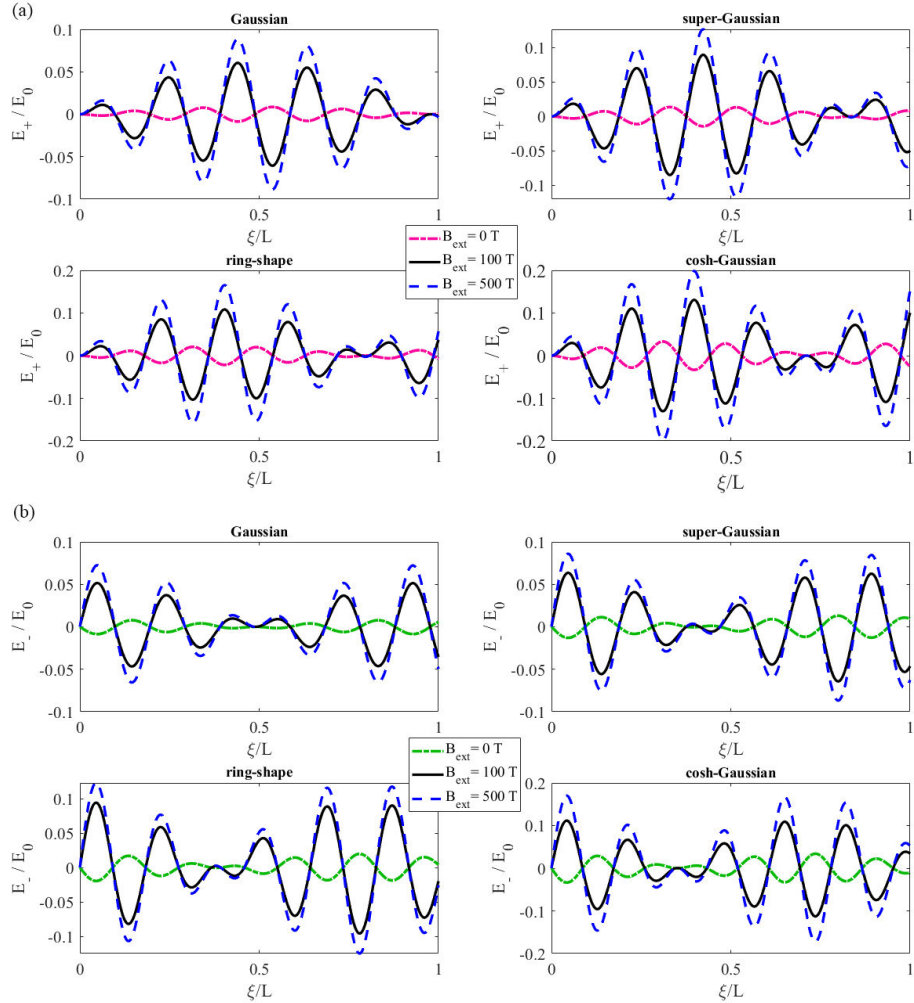

Fig. 5. Variations of magnitude of normalized electric fields as a function of normalized independent variable ($\xi$) over plasma interaction length ($L$) for different external magnetic fields in non-relativistic regime with Gaussian profile and elliptical polarization for (a) forward and (b) wake radiations, $\lambda = 800\ nm, L = 10$mm.

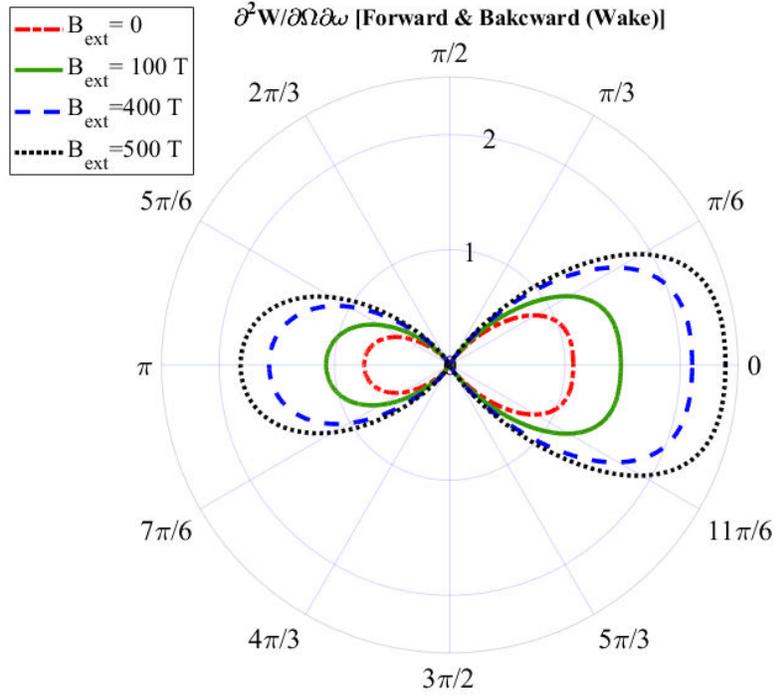

Fig. 6. The impact of external magnetic field on the angular distributions of (a) forward and (b) wake radiations for Gaussian profile and elliptical polarization in relativistic regime, $\lambda = 800\ nm$, $L = 10$mm.

Figure 9 shows the variations of magnitude of normalized electric fields of (a) forward and (b) wake radiations as a function of normalized independent variable ($\xi$) over interaction length for various interaction lengths in the relativistic regime. According to the Figure, as the plasma interaction length increases, both the electric fields of forward wave and the wake amplitudes experience a rise. The extended interaction length allows for prolonged and effective interaction between the laser pulses and plasma, resulting in larger amplitudes for generation of the forward and wake radiation. Despite the increase in electric field amplitudes, the directivity of both fields remains unchanged, and waves exhibit compression. The prolonged interaction length enables a greater number of electrons to participate in the interaction with laser pulses, enhancing the efficiency of energy transfer from the laser pulses to plasma. Therefore, there will be a corresponding increase in the speed of plasma electrons. This enhancement, coupled with increase in plasma frequency and electron mobility, establishes an amended resonance condition, leading to an augmentation of the non-zero drift speed of electrons. The larger drift speed, induces a higher nonlinear electron current density. In other words, the improved resonance condition causes the efficient coupling between the laser pulses and plasma, leading to enhanced fields. However, certain instabilities in the laser-plasma system, such as the stimulated Raman scattering (SRS) and stimulated Brillouin scattering (SBS), lead to damping of certain frequencies.

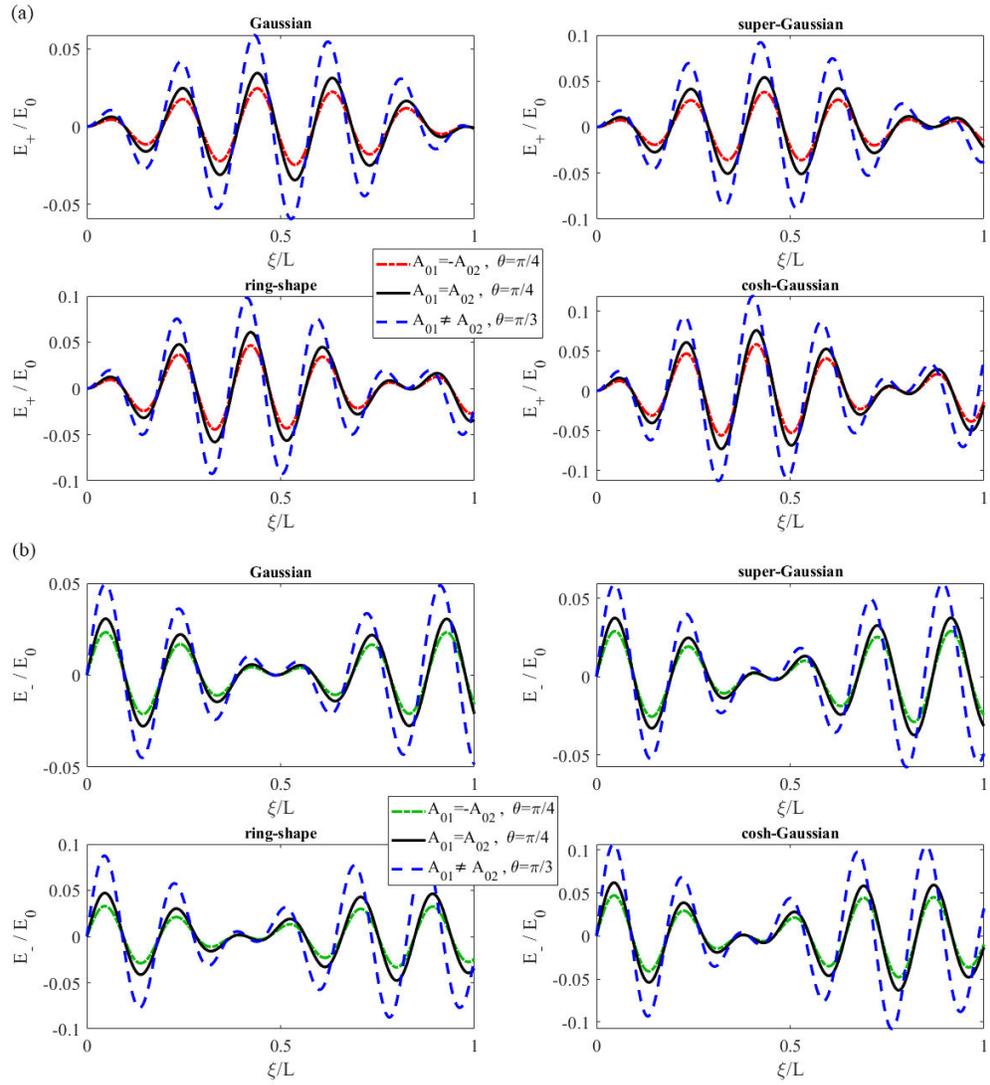

Fig. 7. Role of various polarization states of laser pulses on the magnitude of normalized electric fields of (a) forward and (b) wake radiations for different laser profiles in non-relativistic regime, $\lambda = 800\ nm$, $L = 10$mm and $B_{ext} = 100\ T$.

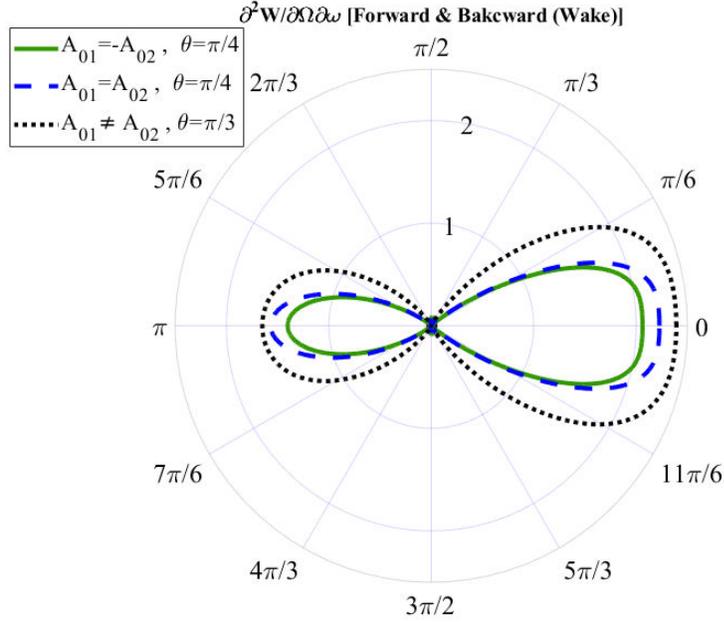

Fig. 8. Plots of angular distributions of (a) forward and (b) wake radiations for different polarization states of laser pulses with Gaussian profile in relativistic regime, $\lambda = 800\ nm$, $L = 10\text{mm}$ and $B_{ext} = 100\ T$.

The impact of interaction length on the angular distributions of the wake and forward waves for Gaussian profile in a relativistic regime is presented in Fig 10. The Figure indicates that as the interaction length increases, despite enhancements of the wake and forward-directed THz wave, the radiation cone angle is decreased. Changes in plasma interaction length introduces notable variations in plasma density and plasma electrons group velocity. An increase in the number of collisions within the plasma environment results in a substantial rise in electron collision frequency. In such a scenario, transfer of energy and momentum from the laser fields to plasma species experience a reduction, leading to a drop in plasma electrons induced nonlinear current density. The drop in current density has a discernible impact on the generated radiation patterns. Specifically, radiation patterns experience a decrease in angular distribution width in backward and forward directions. The experimental results from both Yi Liu et al. and H. Wang et al. validate the observed outcomes pertaining angular distribution and normalized THz wave electric field in plasma [27, 28].

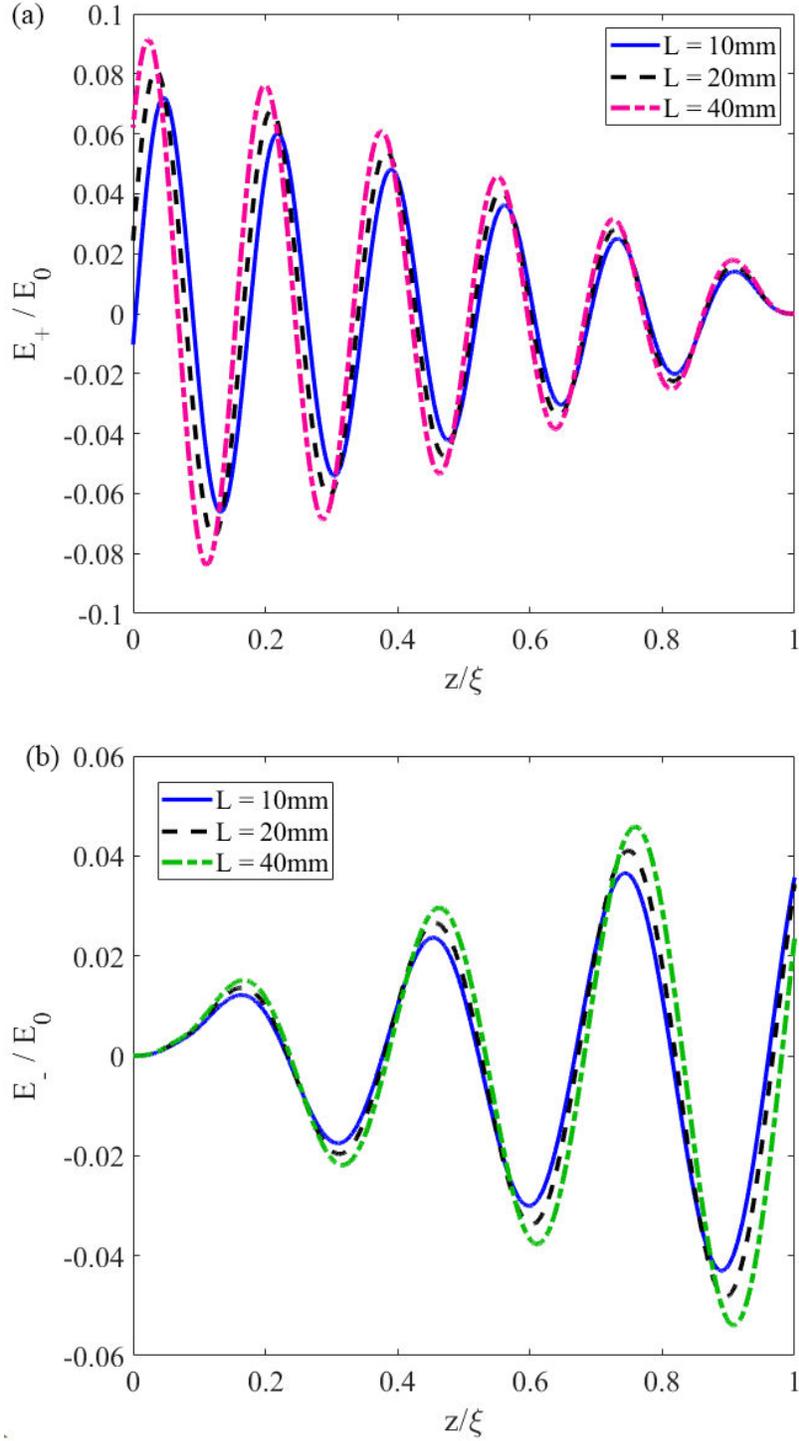

Fig. 9. Variations of magnitude of normalized electric fields as a function of normalized independent variable ($\xi$) over plasma interaction length ($L$) for different plasma interaction lengths and elliptical polarization in relativistic regime for (a) forward and (b) wake radiations, $\lambda = 800\ nm$, $B_{ext} = 100\ T$.

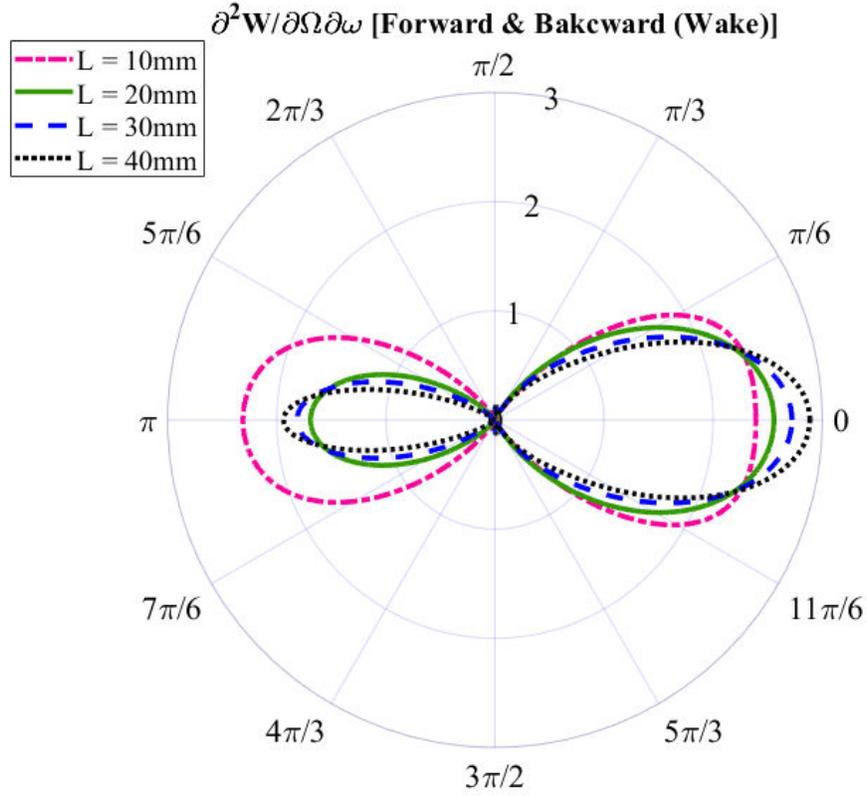

Fig. 10. The impact of interaction length ($L$) on the angular distributions of (a) forward and (b) wake radiations for Gaussian profile and elliptical polarization in relativistic regime, $\lambda = 800\ nm$, $B_{ext} = 100\ T$.

The wake and forward THz wave power as a function of normalized independent variable ($\xi$) over interaction length for Gaussian profile in relativistic and non-relativistic regimes are illustrated in Fig. 11. This Figure provides a perspective on the effectiveness of laser pulse interaction with plasma. The plot facilitates the identification of optimal conditions for the wavefield, as well as forward wave generation. As depicted in the Figure, within the relativistic regime, both waves exhibit a notable power increase. The observed growth can be attributed to the increase in efficiency of the energy transfer from the laser pulses to plasma electrons. In the relativistic regime, the rate of interaction between the laser pulses and plasma electrons intensifies, fostering a robust coupling. The strengthen in coupling, along with increase in plasma frequency and enhanced electron mobility, establishes an improved resonance condition. This sets in the growth of the non-zero drift velocity of electrons and a subsequent increase in the electric fields, leading to enhancement of the generated power of the wake and forward waves.

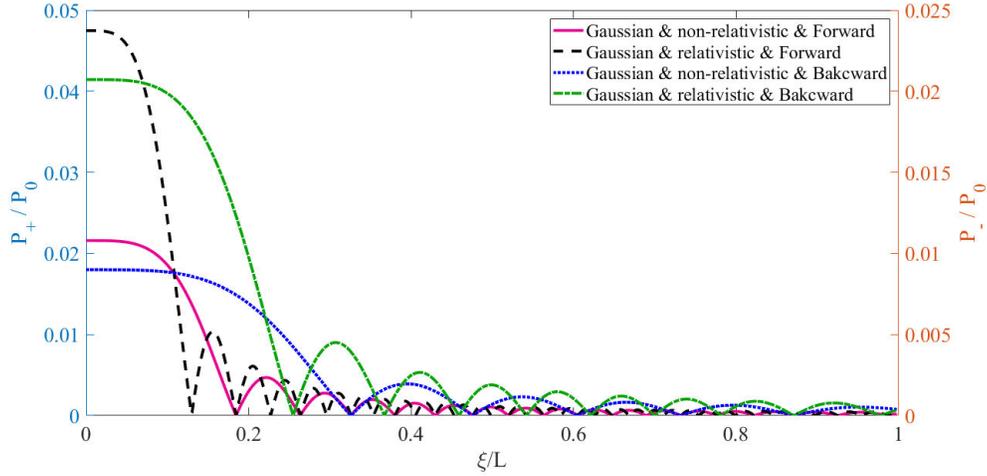

Fig. 11. Variations of magnitude of normalized power of forward and wake wave as a function of normalized independent variable ($\xi$) over plasma interaction length ($L$) for Gaussian profile and elliptical polarization in relativistic and non-relativistic regimes, $\lambda = 800\ nm$, $L = 10$mm, $B_{ext} = 100\ T$.

## 4. Conclusions

The interaction of two-color laser pulses with plasma electrons induces transverse nonlinear currents in two dimensions, leading to generation of forward and wake waves. The simultaneous propagation of the fundamental and second harmonic of the laser pulse for various profiles affects the spectral width and peak frequency of the radiation fields in both forward and backward directions. The analysis included relativistic, as well as non-relativistic regimes. The study indicates that there is a mutual dependency between the electric fields of the wake and the forward wave in the non-relativistic regime of laser-plasma interaction. However, in the relativistic regime, the interaction dynamics undergo a shift, causing waves act independent of each other. Furthermore, the study examined the electric field dynamics, considering the distinctive features and the impact of various parameters in both relativistic and non-relativistic regimes. The time integration of the total energy per frequency per solid angle ($\partial^2 W/\partial\Omega\partial\omega$) was established across various laser profiles. The time integration parameter plays a pivotal role in the far-field emission and determination of radiation patterns along with directionality features. The effect of different parameters such as the number of laser pulse harmonics, interaction length, external magnetic field strength, and the polarization of incident laser pulses on radiation patterns has been verified. The analytical and numerical framework presented in this study introduces a versatile scheme that can be effectively applied for controlling particles within the longitudinal electric field of forward and wake in the THz zone.

### Acknowledgment

This work is based on research funded by Iran National Science Foundation (INSF) under project No. 4021340. The authors would like to express their gratitude towards INSF.

### Disclosures:

The authors declare no conflicts of interest.

**Data availability:**

Data underlying the results presented in this paper are not publicly available at this time but may be obtained from the authors upon reasonable request.

**References**

1. M. Gezimati, Gh. Singh, "Advances in terahertz technology for cancer detection applications," Opt Quantum Electron, 55, 2, 151, (2022), DOI:10.1007/s11082-022-04340-0.
2. A. Abina, U. Puc, M. Jazbinšek, A. Zidanšek, "Analytical Gas Sensing in the Terahertz Spectral Range," Micromachines, 14, 11, 1987, (2023), DOI: 10.3390/mi14111987.
3. T. Zhou, L. Li, Y. Wang, Sh. Zhao, M. Liu, J. Zhu, W. Li, Zh. Lin, J. Li, B. Sun, Q. Huang, G. Zhang, Ch. Zou, "Multifield-Modulated Spintronic Terahertz Emitter Based on a Vanadium Dioxide Phase Transition," ACS Appl Mater Interfaces, 16, 11, 13997-14005, (2024), DOI: 10.1021/acsami.3c19488.
4. Y. Peng, B. Xu, Sh. Zhou, Zh. Sun, H. Xiao, J. Zhao, Y. Zhu, X. Ch. Zhang, D. M. Mittleman, S. Zhuang, "Experimental measurement of the wake field in a plasma filament created by a single-color ultrafast laser pulse," Phys. Rev. E, 102, 6-1, 063211, (2020), DOI: 10.1103/PhysRevE.102.063211.
5. Q. Wang, Y. Chen, J. Mao, F. Yang, N. Wang, "Meta-surface-Assisted Terahertz Sensing," Sensors, 23, 13, 5902, (2023), DOI: 10.3390/s23135902.
6. V. Khudiakov, A. Pukhov, "Optimized laser-assisted electron injection into a quasilinear plasma wakefield," Physics Review E, 105, 3, 035201, (2022), DOI: 10.1103/PhysRevE.105.035201.
7. A. Grigoriadis, G. Andrianaki, I. Tazes, V. Dimitriou, M. Tatarakis, E. P. Benis, N. A. Papadogiannis, "Efficient plasma electron accelerator driven by linearly chirped multi-10-TW laser pulses," Scientific Reports, 20;13(1):2918, (2023), DOI: 10.1038/s41598-023-28755-1.
8. M. Laabs, N. Neumann, B. Green, N. Awari, J. Deinert, S. Kovalev, D. Plettemeier, M. Gensch, "On-chip THz spectrometer for bunch compression fingerprinting at fourth-generation light sources," Journal of Synchrotron Radiation, 1;25:1509-1513, (2018), DOI: 10.1107/S1600577518010184.
9. X. L. Zhu, M. Chen, S. M. Weng, T. P. Yu, W. M. Wang, F. He, Zh. M. Sheng, P. McKenna, D. A. Jaroszynski, J. Zhang, "Extremely brilliant GeV γ-rays from a two-stage laser-plasma accelerator," Science Advances, 9;6(22): eaaz7240, (2020), DOI: 10.1126/sciadv. aaz7240.
10. A. Curcio, V. Dolci, S. Lupi, M. Petrarca, "Terahertz-based retrieval of the spectral phase and amplitude of ultrashort laser pulses," Optics Letters, 15;43(4):783-786, (2018), DOI: 10.1364/OL.43.000783.
11. L. Guiramand, X. Ropagnol, F. Blanchard, "Time-frequency analysis of two-photon absorption effect during optical rectification in a ZnTe crystal pumped at 1.024 μm," Optics Letters, 15;46(24):6047-6050, (2021), DOI: 10.1364/OL.441231.
12. J. Ferri, X. Davoine, S. Fourmaux, J. C. Kieffer, S. Corde, K. T. Phuoc, A. Lifschitz, "Effect of experimental laser imperfections on laser wakefield acceleration and betatron source," Scientific Reports, 6:27846, (2016), DOI: 10.1038/srep27846.
13. X. Yang, E. Brunetti, D. Reboredo Gil, G. H. Welsh, F.Y. Li, S. Cipiccia, B. Ersfeld, D. W. Grant, P. A. Grant, M. R. Islam, M. P. Tooley, G. Vieux, S. M. Wiggins, Z. M. Sheng, D. A. Jaroszynski, "Three electron beams from a laser-plasma wakefield accelerator and the energy apportioning question," Scientific Reports, 7:43910, (2017), DOI: 10.1038/srep43910.
14. X. Yang, E. Brunetti, D. A. Jaroszynski, "High-energy coherent terahertz radiation emitted by wide-angle electron beams from a laser-wakefield accelerator," New J. Phys. 20, 043046, (2018), DOI: 10.1088/1367-2630/aab74d.
15. Y. Peng, B. Xu, Sh. Zhou, Zh. Sun, H. Xiao, J. Zhao, Y. Zhu, Xi-Ch Zhang, D. M. Mittleman, S. Zhuang, "Experimental measurement of the wake field in a plasma filament created by a single-color ultrafast laser pulse," Physical Review E 102, 063211 (2020), DOI: 10.1103/PhysRevE.102.063211.
16. M. T. Hibberd, A. L. Healy, D. S. Lake, V. Georgiadis, E. J. H. Smith, O. J. Finlay, T. H. Pacey, J. K. Jones, Y. Saveliev, D. A. Walsh, E. W. Snedden, R. B. Appleby, G. Burt, D. M. Graham, S. P. Jamison, "Acceleration of relativistic beams using laser-generated terahertz pulses," Nature Photonics, (2020), DOI: 10.1038/s41566-020-0674-1.
17. A. Pukhov, A. Golovanov, I. Kostyukov, "Efficient Narrow Band Terahertz Radiation from Electrostatic Wakefields in Nonuniform Plasmas," Physical Review Letters, 127, 175001, (2021), DOI: 10.1103/PhysRevLett.127.175001.
18. C. Tailliez, X. Davoine, L. Gremillet, L. Bergé, "Terahertz pulse generation from relativistic laser wakes in axially magnetized plasmas," Physical Review Research, 5, 023143 (2023), DOI: 10.1103/PhysRevResearch.5.023143.
19. V. Sharma, V. Thakur, "Lasers wakefield acceleration in under-dense plasma with ripple plasma density profile," J. Opt, (2023), DOI: 10.1007/s12596-023-01548-5.
20. J. J. van de Wetering, S. M. Hooker, R. Walczak, "Multi-GeV wakefield acceleration in a plasma-modulated plasma accelerator," Physical Review E, 109, 025206, (2024), DOI: 10.1103/PhysRevE.109.025206.

21. E. A. Nanni, W. R. Huang, K. H. Hong, K. Ravi, A. Fallahi, G. Moriena, R. J. D. Miller, F. X. Kärtner, "Terahertz-driven linear electron acceleration," nature communications, 6, 8486, (2015), DOI: 10.1038/ncomms9486.
22. V. Tomkus, V. Girdauskas, J. Dudutis, P. Gečys, V. Stankevič, G. Račiukaitis, I. G. González, D. Guénot, J. B. Svensson, A. Persson, O. Lundh, "Laser wakefield accelerated electron beams and betatron radiation from multi-jet gas targets," Scientific Reports, 10:16807, (2020), DOI: 10.1038/s41598-020-73805-7.
23. T. Pak, M. Rezaei-Pandari, S. B. Kim, G. Lee, D. H. Wi, C. I. Hojbota, M. Mirzaie, H. Kim, J. H. Sung, S. K. Lee, Ch. Kang, K. Y. Kim, "Multi-millijoule terahertz emission from laser-wakefield-accelerated electrons," Light: Science & Applications, 12:37, (2023), DOI: 10.1038/s41377-022-01068-0.
24. Y. Peng, B. Xu, Sh. Zhou, Zh. Sun, H. Xiao, J. Zhao, Y. Zhu, X. Ch. Zhang, D. M. Mittleman, S. Zhuang, "Experimental measurement of the wake field in a plasma filament created by a single-color ultrafast laser pulse," Physical Review E 102, 063211 (2020), DOI: 10.1103/PhysRevE.102.063211.
25. D. Ma, L. Dong, M. Zhang, R. Zhang, Y. Zhao, C. Zhang, L. Zhang, "Terahertz wave generation from two-color laser excited air plasma modulated by bi-chromatic laser fields," IEEE Transactions on Terahertz Science and Technology, 12, 3, (2022), DOI: 10.1109/TTHZ.2022.3149975.
26. C. Tailliez, X. Davoine, A. Debayle, L. Gremillet, L. Berg´, "Terahertz Pulse Generation by Strongly Magnetized, Laser-Created Plasmas," physical Review Letters, 128, 174802 (2022), DOI: 10.1103/PhysRevLett.128.174802.
27. Y. Liu, S. Liu, A. Houard, A. Mysyrowicz, V. T. Tikhonchuk, "Terahertz Radiation from a Longitudinal Electric Field Biased Femtosecond Filament in Air," Chin. Phys. Lett. 37 6 (2020), DOI: 10.3390/s22145231.
28. H. Wang, H. Shangguan, Q. Song, Y. Cai, Q. Lin, X. Lu, Z. Wang, S. Zheng, S. Xu, "Generation and evolution of different terahertz singular beams from long gas-plasma filaments," Optics Express 29 2 (2021), DOI: 10.1364/OE.413483.